\documentclass[doublecol,amsmath,amssymb]{epl2}
\usepackage{amsmath}
\usepackage{cases}
\usepackage{amsfonts}
\usepackage{color}
\usepackage{graphicx}
\usepackage{dcolumn}
\usepackage{bm}
\usepackage{amssymb}
\usepackage{array}
\usepackage{longtable}
\usepackage{hhline}

\begin{document}

\title{Complete spectrum of stochastic master equation for random walks on treelike fractals}
\shorttitle{Complete spectrum of stochastic master equation for  random walks on treelike fractals}

\author{Zhongzhi Zhang\inst{1,2} \footnote{\email{zhangzz@fudan.edu.cn}} \and Bin Wu\inst{1,2} \and Guanrong Chen \inst{3} \footnote{ \email{eegchen@cityu.edu.hk}}}
\shortauthor{Zhongzhi Zhang, Bin Wu, and Guanrong Chen}

 \institute{
  \inst{1} School of Computer Science, Fudan University, Shanghai 200433, China\\
  \inst{2} Shanghai Key Lab of Intelligent Information Processing, Fudan University, Shanghai 200433, China\\
  \inst{3} Department of Electronic Engineering, City University of Hong Kong, Hong Kong SAR, China}

\date{\today}

\begin{abstract}{
We study random walks on a family of treelike regular fractals with a trap fixed on a central node. We obtain all the eigenvalues and their corresponding multiplicities for the associated stochastic master equation, with the eigenvalues being provided through an explicit recursive relation. We also evaluate the smallest eigenvalue and show that its reciprocal is approximately equal to the mean trapping time. We expect that our technique can also be adapted to other regular fractals with treelike structures.}
\end{abstract}



 \maketitle

\section{Introduction}

As a valuable tool for various scientific disciplines, fractals introduced by Mandelbrot~\cite{Ma82} have received considerable attention in many fields in the past few decades~\cite{HaBe87,BeHa00,Fa03,AgViSa09}. The so-called regular fractals constitute an integral family of fractals, including the Sierpinski triangle~\cite{Si1915}, hierarchical lattice fractal~\cite{BeOs79}, Vicsek fractals~\cite{Vi83}, $T$ fractal~\cite{KaRe86,KaRe89}, Peano basin fractal~\cite{BaDeVe09}, to name a few. These simple fractal structures have been a focus of extensive research~\cite{Fa03}. A great advantage of regular fractals is that many problems about them can be treated analytically, presenting a first step in understanding the structural and dynamical proprieties of general fractals.

A fundamental problem in the study of fractals is to find the relations between structural properties of fractals and their dynamics, which remains one of the great challenges in the realm of nonlinear science. As a paradigmatic dynamical process, random walks on fractals play a crucial role in a wide range of applications~\cite{HaBe87,BeHa00}, therefore have been intensely investigated~\cite{CoBeTeVoKl07,GaSoHaMa07,HaRo08,ZhXiZhGaGu09, ZhLiZhWuGu09,TeBeVo09,ZhXiZhLiGu09,HaRo09,ZhWjZhZhGuWa10,WeKlBl10,WuZhCh11}. These investigations have uncovered some nontrivial effects of topological structures of different fractals on the behavior of random walks. Among various random-walk issues~\cite{MeKl04,BuCa05}, the trapping problem on fractals are becoming increasingly important since many phenomena built on fractals can be described by random walks~\cite{AmVaGr00,VeChEbHeKo10}. Particularly, the mean first-passage time (MFPT) to the trap node in a network has been widely studied~\cite{Mo69,KaBa02PRE,KaBa02IJBC,KiCaHaAr08,Ag08,ZhQiZhXiGu09,AgBu09,ZhGaXi10} due to its role as an indicator of the transport efficiency.

In view of the importance of MFPT~\cite{Re01}, a huge amount of effort has been devoted to developing techniques to determine MFPT to a given node, and numerous methods have been proposed, including the generating function~\cite{Mo69}, renormalization group~\cite{KaBa02PRE,Ag08}, back-forward equation~\cite{ZhQiZhXiGu09}, and so on. Most of existing methods are presented for particular networks and thus are applicable only to some specific structures. A universal technique for general networks is still unavailable, with the exception of the spectral approach, through which the MFPT from a node to another can be expressed in terms of the eigenvalues and their corresponding eigenvectors of a related transition matrix~\cite{Lo96}. For the trapping problem, the average trapping time (ATT), i.e., the MFPT to the trap, can be approximated by the inverse of the smallest eigenvalue of an associated matrix~\cite{We1994}. Irrespectively of the generality of the spectral approach, the computational complexity for determining the spectra of generic networks is very high~\cite{Bi93}, which makes the task very difficult and even impossible for large networks. In this context, it is of interest to seek new techniques for evaluating the spectrum of a relevant matrix for some particular networks.

In this paper, we study a random walk problem on a family of treelike regular fractals~\cite{LiWuZh10} with a perfect trap fixed at a central node. The fractals under investigation include both the $T$ fractal~\cite{KaRe86,KaRe89} and the Peano basin fractal~\cite{BaDeVe09} as their particular limiting cases. Using the real-space decimation procedure, we derive an explicit recursive relation for the eigenvalues of a matrix associated with the trapping issue. We then continue to determine the full spectrum of eigenvalues and their multiplicities of the relevant matrix. In the end, we derive a recursion relation for the smallest eigenvalues at two consecutive iterations of the fractals, based on which we  obtain an approximate value of the final smallest eigenvalue and then further evaluate the ATT.

\begin{figure}
\begin{center}
\includegraphics[width=0.80\linewidth]{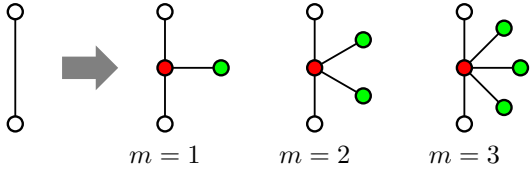}
\caption{ (Color online) Construction of the
fractals. The next generation is obtained through
replacing each edge by the clusters on the right-hand side of the arrow.}
\label{cons}
\end{center}
\end{figure}

\section{Constructions and properties of regular fractals\label{model}}

\begin{figure}
\begin{center}
\includegraphics[width=0.80\linewidth]{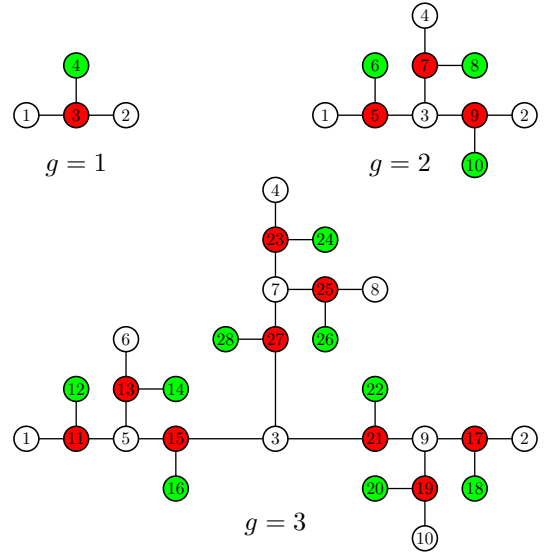}
\end{center}
\caption[kurzform]{ (Color online) Iterative process for a
special fractal corresponding to $m=1$.} \label{network}
\end{figure}

The fractals under consideration are constructed in the following iterative way~\cite{LiWuZh10}. Let $F_{g}$ ($g \ge 0$) denote the fractals after $g$ iterations. Initially ($g=0$), $F_{0}$ consists of an edge linking two nodes. In each subsequent iteration $g \ge 1$, $F_{g}$ is obtained from $F_{g-1}$ by performing the following operations on each existing edge as illustrated by Fig.~\ref{cons}: replace the edge by a path two-links long, with the two end points of the path being the same end points of the original edge; then, create $m$ (an natural number) new nodes and attach them individually  to the middle node of the path. Figure~\ref{network} describes the first three steps of the iterative construction process for a particular fractal corresponding to $m=1$. The family of fractals, controlled by parameter $m$, includes several well-known fractals as its limiting cases: When $m =1$, it reduces to the $T$ fractal~\cite{KaRe86,KaRe89}; when $m = 2$, it recovers the Peano basin fractal~\cite{BaDeVe09}.

\begin{figure}
\begin{center}
\includegraphics[width=0.70\linewidth]{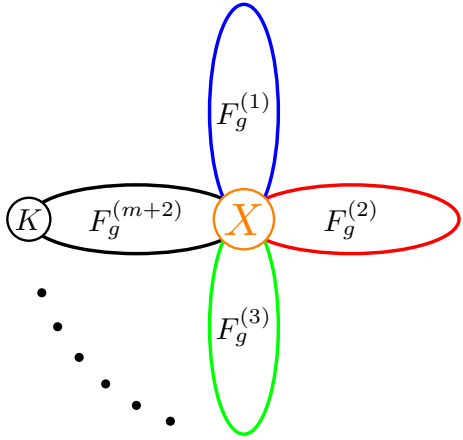}
\end{center}
\caption[kurzform]{(Color online) Second construction method
of the fractals highlighting their self-similarity. We can obtain $F_{g+1}$ by joining $m+2$ copies of $F_{g}$, represented by $F_{g}^{(1)}$, $F_{g}^{(2)}$, $\ldots$,
$F_{g}^{(m+1)}$, and $F_{g}^{(m+2)}$. $X$ is the innermost
node, while $K$ is an outermost node.}\label{Const2}
\end{figure}

The studied fractals have an obvious self-similar structure, which can be seen from their alternative construction approach. If we call the central node (node $3$ in Fig.~\ref{network}, for example) the innermost node and those nodes farthest from the central node the outermost nodes, then the fractals can also be built as follows (see Fig.~\ref{Const2}): Given the generation $g$, we can obtain $F_{g+1}$ by merging $m+2$ copies of $F_{g}$ with the $m+2$ outermost nodes in separate replicas being coalesced into a single new node, i.e., the innermost node in $F_{g+1}$.

It is easy to derive that the total numbers of edges and nodes in $F_{g}$ are
$E_{g}=(m+2)^{g}$ and $N_{g}=(m+2)^{g}+1$, respectively. Some other properties of the fractals can also be determined~\cite{LiWuZh10}. Their fractal dimension is $d_{\rm f}=\ln (m+2)/\ln2$, while their random-walk dimension is $d_{\rm w}=\ln[2(m+2)]/\ln2$. Thus, their spectral dimension is $d_{\rm s}=2d_{\rm f}/d_{\rm w}=2\ln (m+2)/\ln[2(m+2)]$.

\section{Random walks with a trap positioned on the innermost node}

In this section, we study the discrete-time unbiased random walks taking place on the fractals $F_{g}$. In the process of the classical random walks, at each time step, the particle (walker) starts from its current location and moves to any of its nearest-neighboring nodes with the same probability equal to the reciprocal of the degree of the current node.

Now, consider the probability distribution function $\rho_j(t)$, which is the probability for the walker being on node $j$ at time $t$. If the starting point for the walker is $s$, then the stochastic master equation governing the evolution of $\rho_j(t)$ is~\cite{HoGrWe85,BaKl98}
\begin{equation}\label{SME}
\frac{{\rm d} \rho_j(t)}{{\rm d} t}=-\sum_{i=1}^{N_g}P_{ij} \rho_i(t)
\end{equation}
with the initial condition ${\rm d} \rho_j(t=0)=\delta_{j,s}$.

In Eq.~(\ref{SME}), if $i$ is not equal to $j$, $P_{ij}$ is the jumping probability from node $i$ to node $j$, but $P_{ij}=-1$ otherwise ($i=j$). Note that we can consider $P_{ij}$ as an entry of matrix $P_g$, called the transition matrix, which suggests that the powerful methods of the spectral theory for matrices~\cite{CvDoSa95} can be applied.

\subsection{Eigenvalues of the stochastic master equation}

We focus on a trapping problem which is a particular case of random
walks on $F_{g}$ with a trap fixed on the central node
of the fractals. For this case, the transition matrix is a submatrix of
$P_g$, which is obtained from $P_g$ by deleting the row and column corresponding to the trap node. In the case without any confusion, we also use $P_g$ to represent the transition matrix of the trapping problem. Below we will use the decimation method~\cite{DoAlBeKa83,CoKa92,BeTuKo10} to determine the complete spectrum of eigenvalues and their degeneracies, and demonstrate that the reciprocal of the smallest eigenvalue has the same scaling as that of the ATT. Notice that previous works have studied analytically the Laplacian spectra for different treelike graphs, e.g., dendrimers~\cite{CaCh97}, Vicsek fractals~\cite{BlJuKoFe03,BlFeJuKo04}, and related structures, say Husimi-cacti~\cite{GaBl07}. Here we concentrate on the normalized Laplacian matrix, since for an irregular graph, its Laplacian matrix and normalized Laplacian matrix behave quite differently, and the latter seems to be more natural~\cite{ChDaHaLiPaSt04}.

The decimation method has an advantage that one can solve the eigenvalue equation based on a similar problem from the previous iteration. 
Let $\alpha$ denote the set of nodes belonging to $F_{g}$, and $\beta$ the set of nodes generated at iteration $g+1$. Then, one can write the eigenvalue problem for the transition matrix $P_{g+1}$ in the following block form:
\begin{equation}\label{T1}
\left[\begin{array}{cccc}
P_{\alpha,\alpha} & P_{\alpha, \beta} \\
P_{\beta,\alpha} & P_{\beta, \beta}
\end{array}
\right]
\left[\begin{array}{cccc}
 u_{\alpha} \\
 u_{\beta}
\end{array}
\right]={\lambda}_{g+1} \left[\begin{array}{cccc}
 u_{\alpha} \\
 u_{\beta}
\end{array}
\right],
\end{equation}
where $P_{\alpha,\alpha}$ is the identity matrix since there are no transitions between the original nodes, $P_{\beta, \beta}$ is
block diagonal with each block being the same $(m+1)\times(m+1)$ matrix of the form
\begin{equation}\label{T2}
B=\left[\begin{array}{ccccc}
 1 & -\frac{1}{m+2} & -\frac{1}{m+2} & \cdots & -\frac{1}{m+2} \\
 -1 & 1 & 0 & \cdots & 0 \\
 -1 & 0 & 1 & \cdots & 0 \\
 \vdots & \vdots & \vdots & \ddots & \vdots \\
 -1 & 0 & 0 & \cdots & 1
\end{array}
\right],
\end{equation}
and $\lambda_{g+1}$ is an eigenvalue of $P_{g+1}$.

Equation~(\ref{T1}) can be rewritten as two equations:
\begin{equation}\label{T3}
P_{\alpha,\alpha}u_{\alpha}+P_{\alpha,
\beta}u_{\beta}={\lambda}_{g+1}u_{\alpha},
\end{equation}
\begin{equation}\label{T4}
P_{\beta,\alpha}u_{\alpha}+P_{\beta,
\beta}u_{\beta}={\lambda}_{g+1}u_{\beta}.
\end{equation}
From Eq.~(\ref{T4}), one obtains
\begin{equation}\label{T5}
u_{\beta}=({\lambda}_{g+1}-P_{\beta,
\beta})^{-1}P_{\beta,\alpha}u_{\alpha}\,,
\end{equation}
provided that the concerned matrix is invertible. Inserting Eq.~(\ref{T5})
into Eq.~(\ref{T3}) leads to
\begin{equation}\label{T6}
[P_{\alpha,\alpha}+P_{\alpha, \beta}({\lambda}_{g+1}-P_{\beta,
\beta})^{-1}P_{\beta,\alpha}]u_{\alpha}={\lambda}_{g+1}u_{\alpha}.
\end{equation}

Thus, the problem of determining the eigenvalue ${\lambda}_{g+1}$
for matrix $P_{g+1}$ of order $(m+2)^{g+1} \times (m+2)^{g+1}$ is
reduced to computing the eigenvalue problem of a matrix with a smaller
order of $(m+2)^{g} \times (m+2)^{g}$.

Let $Q_{g}=[P_{\alpha,\alpha}+P_{\alpha,
\beta}({\lambda}_{g+1}-P_{\beta, \beta})^{-1}P_{\beta,\alpha}]$.
Next, we will show that matrix $Q_{g}$ can be related to matrix
$P_{g}$, and thus the eigenvalue ${\lambda}_{g+1}$ for matrix
$P_{g+1}$ can be expressed in terms of eigenvalue
${\lambda}_{g}$ for matrix $P_{g}$. To this end, we define
\begin{equation}\label{T7}
R_{g}=(x+y)I_{g}-yP_{g},
\end{equation}
where $I_{g}$ is the identity matrix of the same order as $P_{g}$,
\begin{equation}\label{T8}
x=1+\frac{\lambda_{g+1}-1}{(m+2)(\lambda_{g+1}-2)\lambda_{g+1}+2},
\end{equation}
and
\begin{equation}\label{T9}
y=x-1=\frac{\lambda_{g+1}-1}{(m+2)(\lambda_{g+1}-2)\lambda_{g+1}+2}.
\end{equation}
Equation~(\ref{T7}) indicates that the eigenvalues of $R_{g}$ are related to those of $P_{g}$. In the Appendix, we prove that $Q_{g}=R_{g}$, thus the eigenvalues of $P_{g+1}$ can be obtained through determining the spectrum of $P_{g}$.

Substituting $R_{g}$ for $Q_{g}$, Eq.~(\ref{T6}) becomes
\begin{eqnarray}\label{T10}
Q_{g}u_{\alpha}&=&(x+y)I_g u_{\alpha}-y P_g u_{\alpha}\nonumber\\
&=&{\lambda}_{g+1}u_{\alpha},
\end{eqnarray}
which means
\begin{equation}\label{T11}
(x+y-\lambda_{g+1})u_{\alpha}=y P_{g}u_{\alpha},
\end{equation} that is,
\begin{equation}\label{T12}
P_g u_{\alpha}=\frac{x+y-\lambda_{g+1}}{y}u_{\alpha}.
\end{equation}
If $\lambda_{g}$ denotes the eigenvalue of $P_{g}$ corresponding to the eigenvector $u_{\alpha}$,
 Eq.~(\ref{T12}) implies
\begin{equation}\label{T13}
\lambda_{g}=\frac{x+y-\lambda_{g+1}}{y}.
\end{equation}
Plugging  Eqs.~(\ref{T8}) and~(\ref{T9}) in  Eq.~(\ref{T13}), one can see that the
eigenvalue ${\lambda_{g+1}}$ for transition matrix $P_{g+1}$ is related
to that of $P_{g}$ via
\begin{equation}\label{T14}
\lambda_{g}=-(m+2)(\lambda_{g+1}-2)\lambda_{g+1},
\end{equation}
which can be recast as
\begin{equation}\label{T15}
(m+2)\left(\lambda_{g+1}\right)^2-2(m+2)\lambda_{g+1}+\lambda_{g}=0.
\end{equation}
Solving Eq.~(\ref{T15}) gives its two roots
\begin{equation}\label{T16}
\lambda_{g+1}=1 \pm \sqrt{1-\frac{\lambda_{g}}{m+2}}\,,
\end{equation}
which provide two eigenvalues of $P_{g+1}$.

By calculating the eigenvalues numerically, one can find some important properties of the spectra. First, all eigenvalues appearing at a given generation $g_i$ always exist at the next generation $g_{i}+1$. Second, all new eigenvalues at generation $g_{i}+1$ are just those generated via Eq.~(\ref{T16}) by substituting $\lambda_{g}$ with $\lambda_{g_i}$ that are newly added to generation $g_{i}$. That is, all eigenvalues can be obtained by the recurrence relation Eq.~(\ref{T16}). Then, the next step is to determine the degeneracies of the eigenvalues, based on the two characteristics.

\subsection{Multiplicities of eigenvalues}

Let $N_g (\lambda)$ denote the multiplicity of eigenvalue $\lambda$ of the transition matrix $P_g$. Since each eigenvalue in the fractals can be derived from eigenvalue 1, we first determine the number of  eigenvalue 1 of $P_g$. For this purpose, let $r(M)$ be the rank of matrix $M$. Then, the number
of eigenvalue of $P_g$ is
\begin{equation}\label{N0}
N_g (\lambda=1)= (m+2)^g-r(P_g-1\times I_g)\,.
\end{equation}

For the two limiting cases of $g=1$ and $g=2$, we can easily obtain $N_1(\lambda=1)$ and $N_2(\lambda=1)$.
When $g=1$, $P_g$ is the identity matrix of order $(m+2) \times (m+2)$, so $N_1 (\lambda=1)=m+2$.
When $g=2$, $P_g-I_g$ is block diagonal, with each of its $m+2$ blocks being the same
$(m+2) \times (m+2)$ matrix in the form of
\begin{equation}\label{N1}
\left[\begin{array}{ccccc}
 0 & -\frac{1}{m+2} & -\frac{1}{m+2} & \cdots & -\frac{1}{m+2} \\
 -1 & 0 & 0 & \cdots & 0 \\
 -1 & 0 & 0 & \cdots & 0 \\
 \vdots & \vdots & \vdots & \ddots & \vdots \\
 -1 & 0 & 0 & \cdots & 0
\end{array}
\right].
\end{equation}
Obviously, the rank of each block is 2. Thus, we have $N_2 (\lambda=1)=(m+2)^2-2(m+2)=m(m+2)$.

In order to determine $N_g (\lambda=1)$ for the case of $g>2$, we define $B_g$ as
the transition matrix of $F_g$ with the trap located at an outmost node, e.g., node $K$ in Fig.~\ref{Const2}, and define $A_g=B_g-I_g$.

We next show that it is possible to determine exactly the multiplicity of eigenvalue 1 for the fractals at every iteration $g>2$.

Note that for $g \geq 2$, $P_{g+1}-I_{g+1}$ has the following form:
\begin{equation}\label{N2}
P_{g+1}-I_{g+1}=\left[\begin{array}{ccccc}
 A_{g} & 0 & 0 & \cdots & 0 \\
 0 & A_{g} & 0 & \cdots & 0 \\
 0 & 0 & A_{g} & \cdots & 0 \\
 \vdots & \vdots & \vdots & \ddots & \vdots \\
 0 & 0 & 0 & \cdots & A_{g}
\end{array}
\right],
\end{equation}
where $A_g$ $(g \geq 2)$ satisfies
\begin{equation}\label{N3}
A_{g}=\left[
\begin{array}{ccccc}
 A_{g-1} & 0 & 0 & \cdots & -v_1 \\
 0 & A_{g-1} & 0 & \cdots & -v_2 \\
 0 & 0 & A_{g-1} & \cdots & -v_3 \\
 \vdots & \vdots & \vdots & \ddots & \vdots \\
 -v_1^{\top} & -v_2^{\top} &-v_3^{\top} & \cdots & A_{g-1}
\end{array}
\right],
\end{equation}
in which each $v_i$ ($i=1,2,\cdots,m+2$) is a matrix of order
$(m+2)^g \times (m+2)^g$, which has $(m+2)^{2g}-1$ elements 0 and only
one nonzero element $1/(m+2)$ describing the transition probability
from one node in a replica of $F_{g}^{(i)}$ to the innermost node
being joined together to form $F_{g+1}$ (see Fig.~\ref{Const2}); and
the superscript $\top$ of a matrix represents its transpose.

By construction, for any node $u$ in $F_{g}$($g \geq 2$) that is
directly connected to the center node, it must have a neighbor $h$
with degree 1. Thus, in matrix $A_{g}$, there is only
one nonzero element for row $h$ and for column $h$, respectively, namely, $(h,u)=-1$
and $(u,h)=-1/(m+2)$. Hence, we can use elementary operations of
a matrix to eliminate all  nonzero element $-1/(m+2)$ on the last
row and the last column of $A_{g}$. Then, we have
$r(A_{g})=r(P_{g}-I_{g})$, which implies that
$r(P_{g+1}-I_{g+1})=(m+2)r(P_g-I_g)$. This, together with
$r(P_{2}-I_{2})=2(m+2)$, leads to $r(P_{g}-I_{g})=2(m+2)^{g-1}$.
Combining the above-obtained results, the degeneracy of
eigenvalue 1 is
\begin{equation}
N_g(\lambda=1)=\begin{cases}
m+2, &g=1, \\
m(m+2)^{g-1}, &g \geqslant 2.
\end{cases}
\end{equation}

We proceed to determine the multiplicities of other eigenvalues. For
the convenience of description, on the basis of Eq.~(\ref{T14}), we
define a set
\begin{equation}
\Omega_g=\begin{cases}
\{1\}, &g=1,\\
\{x|(m+2)(2x-x^2)\in \Omega_{g-1}\}, &g>1.
\end{cases}
\end{equation}
It is easy to verify that the cordiality (number of elements of a
set) of $\Omega_g$ is $|\Omega_g|=2^{g-1}$. On the other hand, from
the above discussion, we can conclude that all the eigenvalues of $P_g$
belong to the  set $\{\Omega_1, \Omega_2, \cdots, \Omega_g\}$. Moreover,
extensive numerical computation on the eigenvalues of the studied
fractals shows that the multiplicities of eigenvalues for
the transmission matrix of the fractals satisfy $N_g(\lambda \in \Omega_{i})=N_{g-1}(\lambda \in
\Omega_{i-1})$. Using this property, we can obtain the degeneracy of
each eigenvalue $\lambda$ in $\Omega_i$ as
\begin{equation}
N_g(\lambda \in \Omega_i)=\begin{cases}
m+2, &i=g,\\
m(m+2)^{g-i}, &i < g.
\end{cases}
\end{equation}
Taking into account the degeneracies so obtained, the
number of eigenvalues of $P_g$ is found to be
 \begin{eqnarray}\label{N6}
\sum_{i=1}^g N_g(\lambda \in \Omega_i)|\Omega_i|&=&\sum_{i=1}^{g-1}
m(m+2)^{g-i}2^{i-1}+(m+2)2^{g-1}\nonumber \\
&=&(m+2)^g,
\end{eqnarray}
which indicates that we have found all the eigenvalues of $P_g$.

\subsection{Smallest eigenvalue and average trapping time}

The probability distribution function $\rho(t)$ considered above is very important since it describes the evolution of the system. Many quantities of interest are encoded in this underlying probability distribution function. For example, in the limit of large system sizes, the ATT denoted by $T_g$ for the trapping problem on $F_g$ is asymptotically equal to the inverse of the smallest nonzero eigenvalue of $P_g$, denoted by $z_g$, that is, $T_g \simeq 1/z_g$. Here, $T_g$ is defined as the average of trapping time $T_g(v)$ of node $v$ over all nodes except the innermost trap node. Conceptually, $T_g(v)$ is the expected time for a walker starting from node $v$ to reach the trap for the first time.

According to the recurrence relation given by Eq.~(\ref{T16}) that dominates the eigenvalues of the transmission matrix for the fractals at two consecutive iterations, we can evaluate the smallest nonzero eigenvalue $z_g$ of $P_g$. From Eq.~(\ref{T16}), we obtain the following obvious recursive relation:
\begin{equation}\label{S1}
z_{g+1}=1-\sqrt{1-\frac{z_g}{m+2}}.
\end{equation}
Using Taylor's formula, we have
\begin{equation}\label{S2}
z_{g+1}\approx
1-\left(1-\frac{1}{2}\frac{z_g}{m+2}\right)=\frac{z_g}{2(m+2)}\,.
\end{equation}
Considering the initial condition $z_{1}=1$, Eq.~(\ref{S2}) can be solved by induction to yield \begin{equation}\label{S3}
z_g \simeq \frac{1}{[2(m+2)]^{g-1}}\,.
\end{equation}
Then, for large networks,
\begin{equation}\label{S4}
T_g \simeq \frac{1}{z_g}=2^{g-1}(m+2)^{g-1}\,.
\end{equation}

We continue to represent $T_g$ as a function of the network
size $N_{g}$, in order to derive the scaling between these two
quantities. From $N_{g}=(m+2)^{g}+1$, we have
$g=\log_{m+2}(N_{g}-1)$. Hence, for large systems, $T_g$ can be expressed in terms of $N_g$ as
\begin{eqnarray}\label{S5}
\langle T \rangle_{g}\sim (N_{g})^{1+\ln2/\ln(m+2)}=
(N_{g})^{2/d_{\rm s}}\,.
\end{eqnarray}
Thus, for the entire family of the studied fractals, in the limit of large system sizes, the ATT grows as a power-law function of the network size $N_{g}$.
We note that the obtained exponent $2/d_{\rm s}$ perfectly agree with the previously obtained result~\cite{Ag08,HaRo08,LiWuZh10}.

\section{Conclusions}

The so-called discrete-time random-walk problem on a graph can be described by a matrix, therefore the distribution of its eigenvalue spectrum is much of interest. In this paper, we have studied the trapping problem on a class of regular treelike fractals with a deep trap located at the central node. The self-similar structure of these fractals can be characterized analytically in terms of their eigenvalues. We have derived a recursive relationship between eigenvalues of a relevant matrix for random walks on the fractals at different iterations by making use of  renormalization, for which the computational demands are very low. On the basis of the obtained recursion relation, we have characterized the complete set of eigenvalues and further determined their degeneracies. Moreover, we have provided a recursion relation between the smallest nonzero eigenvalues of the fractals at two successive iterations and obtained an approximate expression for the smallest nonzero eigenvalue. Finally, we have evaluated the MFPT to the trap, showing that it is approximately equal to the inverse of the smallest eigenvalue.

\begin{acknowledgments}
This work was supported by the National Natural Science Foundation
of China under Grant No. 61074119 and the Hong Kong Research Grants
Council under the GRF Grant CityU 1114/11E.
\end{acknowledgments}


\section{Proof of the equivalence of matrices $Q_g$ and $R_g$ \label{AppA}}

In order to prove that $Q_g=R_g$, it suffices to show that their corresponding entries, denoted by $(Q_g)_{i,j}$ and $(R_g)_{i,j}$, are equal to each other. Since both $Q_g$ and $R_g$ are related to $P_{g+1}$ (or $P_{g}$), we first determine the entry $(P_{g+1})_{i,j}$ of $P_{g+1}$ before evaluating $(Q_g)_{i,j}$ and $(R_g)_{i,j}$.

By the construction of the fractals, the element $(P_{g+1})_{i,j}$ of the block matrix (see Eq.~(\ref{T1})) can be easily found. According to the properties of fractals, we have $(P_{\alpha,\alpha})_{i,j}=\delta_{i,j}$, where $\delta_{i,j}$ is the delta function defined as $\delta_{i,j}=1$ if $i=j$, and $\delta_{i,j}=0$ otherwise. For $(P_{\alpha,\beta})_{i,j}$ and $(P_{\beta,\alpha})_{i,j}$, we distinguish two cases: if the nodes $i$ and $j$ are directly connected by an edge in $F_{g+1}$, then $(P_{\alpha,\beta})_{i,j}=(P_{\beta,\alpha})_{i,j}=-1/k_i$, where $k_i$ is the degree of node $i$ in $F_{g+1}$; otherwise $(P_{\alpha,\beta})_{i,j}=(P_{\beta,\alpha})_{i,j}=0$. Finally, for the diagonal block matrix $P_{\beta,\beta}$, each entry is given in Eq.~(\ref{T2}).

Notice that $Q_g$ is relevant to the inverse of the matrix $\lambda_{g+1}-P_{\beta,\beta}$.
So we should determine $(\lambda_{g+1}-P_{\beta,\beta})^{-1}$. Recall that $P_{\beta,\beta}$ is block diagonal, then $(\lambda_{g+1}-P_{\beta,\beta})^{-1}$ is also block diagonal with each block being $(\lambda_{g+1}-B)^{-1}$. Since our goal is to calculate $(Q_g)_{i,j}$, as will be shown, it suffices to find the entry of $(\lambda_{g+1}-B)^{-1}$ at its first line and first column, which is denoted by $\Theta$. By definition, we have
\begin{equation}\label{H1}
(\lambda_{g+1}-B)^{-1}=\frac{(\lambda_{g+1}-B)^{*}}{{\rm det}(\lambda_{g+1}-B)},
\end{equation}
where $(\lambda_{g+1}-B)^{*}$ is the complex adjugate matrix of $(\lambda_{g+1}-B)$. It is easy to verify that
\begin{equation}\label{H2}
{\rm det}(\lambda_{g+1}-B)=(\lambda_{g+1}-1)^{m+1}-\frac{m(\lambda_{g+1}-1)^{m-1}}{m+2},
\end{equation}
which yields
\begin{equation}\label{H3}
\Theta=\frac{(\lambda_{g+1}-1)^m}{{\rm det}(\lambda_{g+1}-B)}=\frac{(m+2)(\lambda_{g+1}-1)}{2+(m+2)\lambda_{g+1}(\lambda_{g+1}-2)}.
\end{equation}

We are now in a position to determine $(Q_g)_{i,j}$ and $(R_g)_{i,j}$. We distinguish three cases.

Case I: $i=j$. In this case, it is easy to show that
$(R_{g})_{i,j}=x$ and
\begin{equation}\label{H4}
(Q_{g})_{i,j}=1+k_i\left(-\frac{1}{k_i}\right)\Theta
\left(-\frac{1}{m+2}\right)=x=(R_{g})_{i,j}.
\end{equation}

Case II: $i \neq j$, and at the same time nodes $i$ and $j$ are not
directly linked to each other in  $F_{g}$. In the case, it is
obvious that $(Q_{g})_{i,j}=(R_{g})_{i,j}=0$.

Case III: $i \neq j$ but nodes $i$ and $j$ are directly connected in
$F_{g}$. In this case, $(R_{g})_{i,j}=-y(-1/k_i)=y/k_i$ and
\begin{equation}\label{H5}
(Q_{g})_{i,j}=\left(-\frac{1}{k_i}\right)\Theta\left(-\frac{1}{m+2}\right)=\frac{y}{k_i}=(R_{g})_{i,j}.
\end{equation}

Thus, we have proved that $Q_{g}=R_{g}$.

\end{document}